\documentclass[useAMS,usenatbib]{mn2e} 

\usepackage[english]{babel}
\usepackage{journals}
\usepackage[pdftex]{graphicx,color} 
\usepackage[latin1]{inputenc}
\usepackage{graphics} 
\usepackage{amsfonts} 
\usepackage{amsmath}
\usepackage{multicol} 
\usepackage{layout} 
\usepackage{amssymb}
\usepackage[a4paper,colorlinks=true,pdfstartview=FitV,linkcolor=red,citecolor=blue,urlcolor=magenta]{hyperref}
\usepackage{layouts}
\usepackage{multirow}
\usepackage{float}
\usepackage[caption = false]{subfig}
\usepackage{tabularx}
\usepackage{siunitx}

\title[LRG weak lensing]{Probing galaxy assembly bias with LRG weak lensing observations
}

\author[Niemiec et al.]{\parbox{\textwidth}
{A. Niemiec$^{1}$\thanks{E-mail: \href{mailto:anna.niemiec@lam.fr} {anna.niemiec@lam.fr}}, E. Jullo$^{1}$, A. D. Montero-Dorta$^{2}$, F. Prada$^{3}$,  S. Rodriguez-Torres$^{4}$, E. Perez$^{3}$, A. Klypin$^{5}$, T. Erben$^{6}$, M. Makler$^{7}$, B. Moraes$^{8}$, M. E. S. Pereira$^{9}$,  H. Shan$^{6}$} \\
$^{1}$Aix Marseille Univ, CNRS, LAM, Laboratoire d'Astrophysique de Marseille, Marseille, France \\
$^2$ Departamento de F\'isica Matem\'atica, Instituto de F\'isica, Universidade de S\~ao Paulo, Rua do Mat\~ao 1371, CEP 05508-090, S\~ao Paulo, Brazil  \\
$^3$ Instituto de Astrof\'{\i}sica de Andaluc\'{\i}a (CSIC), Glorieta de la Astronom\'{\i}a, E-18080 Granada, Spain \\
$^4$ Dpto. de F\'{\i}sica Te\'orica, Universidad Aut\'onoma de Madrid, Cantoblanco, E-28049 Madrid, Spain \\
$^5$ Astronomy Department, New Mexico State University, Las Cruces, NM 88003, USA\\   
$^6$ Argelander-Institut f\"ur Astronomie, Auf dem H\"ugel 71, 53121 Bonn, Germany \\
$^7$ Centro Brasileiro de Pesquisas F\'isicas, Rua Dr Xavier Sigaud 150, CEP 22290-180, Rio de Janeiro, RJ, Brazil \\
$^8$ Dept. of Physics and Astronomy, University College London, London, WC1E 6BT, UK \\
$^9$ Brandeis University, 415 South Street, Waltham, MA 02453, USA
}

\begin{document}
\date{}
\maketitle
\label{firstpage}
\pagerange{\pageref{firstpage}--\pageref{lastpage}} \pubyear{2018}

\begin{abstract}
In Montero-Dorta et al. 2017, we show that luminous red galaxies (LRGs) from the SDSS-III Baryon Oscillation Spectroscopic Survey (BOSS) at $z\sim0.55$ can be divided into two groups based on their star formation histories. So-called fast-growing LRGs assemble $80\%$ of their stellar mass at $z\sim5$, whereas slow-growing LRGs reach the same evolutionary state at $z\sim1.5$. 
We further demonstrate that these two subpopulations present significantly different clustering properties on scales of $\sim1 - 30 \mathrm{Mpc}$.
Here, we measure the mean halo mass of each subsample using the galaxy-galaxy lensing technique, in the $\sim190\deg^2$ overlap of the LRG catalogue and the CS82 and CFHTLenS shear catalogues. We show that fast- and slow-growing LRGs have similar lensing profiles, which implies that they live in haloes of similar mass: $\log\left(M_{\rm halo}^{\rm fast}/h^{-1}\mathrm{M}_{\odot}\right) = 12.85^{+0.16}_{-0.26}$ and  $\log\left(M_{\rm halo}^{\rm slow}/h^{-1}\mathrm{M}_{\odot}\right) =12.92^{+0.16}_{-0.22}$. This result, combined with the clustering difference, suggests the existence of galaxy assembly bias, although the effect is too subtle to be definitively proven given the errors on our current weak-lensing measurement. We show that this can soon be achieved  with upcoming surveys like DES.

\end{abstract}


\section{Introduction}

Galaxy clustering has been shown to depend on different galaxy properties such as stellar mass, luminosity, colour or star formation rate \citep[e.g][]{coil2008, guo2013, coil2017}.
Currently, this dependence is theoretically explained through halo occupation distribution or halo abundance matching modeling, as resulting from the variations of two underlying factors, the halo mass and the satellite fraction of the studied galaxy sample \citep{zehavi2011, rodriguez-torres2016}. 

However, results from cosmological simulations show that halo clustering not only depends on halo mass, but also on the accretion history of haloes.  At fixed halo mass, haloes that assemble earlier are found to be more tightly clustered than those that assemble at later times, an effect known as halo assembly bias \citep[see e.g.][]{gao2005, wechsler2006, gao&white2007, wang2011, sunayama2016}. 
On the observational side, some studies have used galaxy distributions as tracers for the halo properties to detect halo assembly bias, but the results are still highly debated \citep{miyatake2016, more2016, zu2017b, busch&white2017, dvornik2017}. A complementary question is whether such dependence of the galaxy clustering signal on a secondary \textit{galaxy} property, not related to halo mass, can be found. This effect, which we refer to as \textit{galaxy assembly bias}, has not been solved either \citep{yang&vandenbosch2006, zentner2014, lin2016}.
If the secondary galaxy property can be shown to correlate with the halo formation history, the galaxy assembly bias would be a manifestation of the halo assembly bias.

In \citet{montero-dorta2017} we measured the star formation histories (SFH) of luminous red galaxies (LRGs).
We found that LRGs can be divided into two different types according to their SFH: fast-growing LRGs assemble $80\%$ of their stellar mass very early on ($z \sim 5$), whereas slow-growing LRGs reach the same evolutionary state at $z \sim 1.5$.
These two different evolutionary paths result in very similar populations of massive quiescent galaxies at $z \sim 0.55$, with no significant difference in their stellar mass distributions.
However, the clustering analysis presented in \citet{montero-dorta2017} shows that the clustering amplitude between the two population differs by around $20\%$ in scales between $\sim 1$ and 30 Mpc, with fast-growing LRGs being more strongly clustered than their slow-growing counterparts. 
This difference in the clustering properties at large scale, combined with a similar stellar mass distribution for the two populations is what motivates the work presented here. 
We compare the measured galaxy-galaxy lensing profiles of the two samples to determine whether the clustering difference can be explained by the two populations living in haloes of different masses, or if it should be attributed to a secondary property \textit{traced} by the SFH, which would therefore be an evidence of galaxy assembly bias.

This letter is organized as follows. In Section \ref{sec:data} we present the slow and fast-growing LRG samples, and the CS82 and CFHTLenS shear catalogues. In Section \ref{sec:lensing} we describe the lensing measurement results. In Section \ref{sec:halo_mass} we describe the halo model used to fit the data and the halo masses obtained. We discuss our results and conclude in Section \ref{sec:conclusion}.
Throughout this letter we assume a flat $\Lambda$CDM cosmology  with $(\Omega_{\rm M}, \Omega_{\Lambda}, h, \sigma_8, w) = (0.307, 0.693, 0.678, 0.823,-1)$ \citep{planck2014}, consistently with \citet{montero-dorta2017}. When relevant, the dependence on $h$ is clearly stated.

\section{Data}
\label{sec:data}

	\subsection{The LRG catalogue}

In this work, we use the same LRG catalogue employed in \citet{montero-dorta2017}. The catalogue was extracted from the BOSS CMASS (for "Constant MASS") sample of the Twelfth Data Release of the SDSS \citep[DR12, ][]{alam2015}. From the parent CMASS sample, we selected galaxies in the redshift range $0.5 < z < 0.6$ to maximize stellar-mass completeness and minimize selection effects, and applied a colour cut $g-i > 2.35$ to remove blue objects from the sample. The resulting sample represents a homogeneous population of quenched massive galaxies with $M_* > 10^{11} \mathrm{M}_{\odot}$ at mean redshift $z = 0.55$, containing 305,741 galaxies over a total area of $9,376 \deg^2$.

In \citet{montero-dorta2017}, the LRG catalog is split into two subsamples based on the SFHs, which are measured using the \textsc{Starlight} code \citep{cid_fernandes2005}. The slow-growing LRG subsample contains galaxies with star formation rate at 3 Gyr galaxy-frame look-back time\footnote{Measured from redshift $z = 0.55$, i.e. $t = 5.5$ Gyr.} greater than 2 $\mathrm{M}_{\odot}\mathrm{yr}^{-1}$. The fast-growing LRG subsample is comprised by galaxies that are already quiescent at this time ($\mathrm{SFR} < 2$ $\mathrm{M}_{\odot}\mathrm{yr}^{-1}$). 
In this work we measure the weak lensing signal for these two samples.

	\subsection{The shear catalogues}
	
We measure the LRG halo masses on the overlap between the BOSS survey and two imaging surveys, CFHT/MegaCam Stripe 82 (CS82) and CFHTLenS.

The CS82 Survey \citep{moraes2014} imaged the $173 \deg^2$ ($129 \deg^2$ after masking) of the SDSS Stripe 82 region in the optical \textit{i}-band down to magnitude $24.1$. The redshift for the galaxies were measured  in \citet{bundy2015} using the BPZ algorithm \citep{benitez2000} on the \textit{ugriz} bands from SDSS data. The galaxy shapes used to construct the shear catalogue were measured by the CS82 collaboration \citep{shan2017} using \texttt{lensfit} \citep{miller2013}.

The CFHTLenS survey \citep{heymans2012} covers $154\deg^2$ ($146.5 \deg^2$ after masking) in the \textit{u*g'r'i'z'} bands, and is based on data from the CFHT Legacy Survey.
The redshifts and galaxy shapes are measured by the CFHTLenS collaboration using  BPZ \citep{hildebrandt2012} and \texttt{lensfit} \citep{miller2013}, respectively.
	
For both catalogues we select galaxies with $\textsc{mask} \leq 1$, $\textsc{star\_flag} = 0$, $\textsc{fitclass} = 0$, $z < 1.3$ and $\textsc{weight} > 0$, as described in \citet{niemiec2017}.
After these cuts, the effective weighted source density $n_{\mathrm{eff}} = \frac{1}{\Omega}\frac{(\sum w_i)^2}{\sum w_i^2}$, where $\Omega$ is the total effective area and $w_i$ the \texttt{lensfit} weights of the galaxies \citep{heymans2012}, is $n_{\mathrm{eff}} = 6.7$ galaxies/arcmin$^2$ for the CS82 catalogue and $n_{\mathrm{eff}} = 10.7$ galaxies/arcmin$^2$ for CFHTLenS. The total overlapping region between our BOSS LRG sample and the shear catalogues covers around $190 \deg^2$ and contains 6,972 LRGs (3,356 fast- and 3,616 slow-growing).

\section{Lensing measurement}
\label{sec:lensing}

Galaxy-galaxy lensing is a powerful tool to measure the projected mass density surrounding galaxies, independently of the type or dynamical state of matter. Since the first detection in \citet{brainerd1996}, it has been successfully used in many studies to constrain the galaxy-halo connection, by allowing the measurement of the mass of dark matter haloes \citep[e.g.][]{velander2014, vanuitert2016, leauthaud2016}. This measurement is statistical in nature, as only the lensing signal stacked over an important number of lens galaxies can be detected. In this work, we compare the mean lensing signal computed for fast- and slow-growing LRGs.

The galaxy-galaxy lensing observable is the excess surface mass density profile $\Delta\Sigma(R)$, expressed in comoving units as a function of the projected distance $R$ to the centre of the stacked lenses. 
This quantity can be expressed in terms of the surface mass density of the lens $\Sigma$ as
\begin{equation}
	\Delta\Sigma(R) = \bar{\Sigma}(< R) - \bar{\Sigma}(R)\mathrm{,}
\end{equation}
where $\bar{\Sigma}(R)$ and $\bar{\Sigma}(<R)$ are the surface density averaged respectively on a circle and on a disk of radius $R$ centered on the lens.
The excess surface mass density profile is related to the tangential shear $\gamma_{\mathrm t}$ as:
\begin{equation}
	\Delta\Sigma(R) = \Sigma_{\mathrm{crit}}\gamma_{\mathrm t}(R)\mathrm{,}
\end{equation}
where $\Sigma_{\mathrm{crit}}$ is the critical surface density, expressed in comoving units as:
\begin{equation}
	\Sigma_{\mathrm{crit}} = \frac{c^2}{4\pi G}\frac{D_{\mathrm s}}{D_{\mathrm l}D_{\mathrm{ls}}}\frac{1}{(1 + z_{\mathrm l})^2}\mathrm{.}
\end{equation}
$D_{\mathrm s}$ and $D_{\mathrm l}$ are the angular diameter distances respectively to the source and to the lens, $D_{\mathrm{ls}}$ is the distance between the lens and the source, and $z_{\mathrm l}$ is the redshift of the lens.

In practice, $\Delta\Sigma(R)$ is measured by stacking the signal of each lens-source pairs in 8 logarithmic radial bins from $0.03$ to $3h^{-1}\mathrm{Mpc}$. To account for the errors on the shape measurement, we need to include the inverse variance weight factor $w_{\mathrm s}$ \citep{heymans2012}. Also a multiplicative calibration factor $m$ needs to be applied to the measured signal in a statistical way \citep{miller2013}, which gives a corrected lensing signal measured as described in equations 17-19 in \citet{niemiec2017}.
To ensure that the sources are in the background compared to the lenses, we use only lens-source pairs with $z_{\mathrm{source}} > z_{\mathrm{lens}} + z_{\mathrm{lens}}^{\mathrm{err}} + z_{\mathrm{source}}^{\mathrm{err}}$. We compute covariance matrices with a block bootstrap on the data: we divide the field in blocks, and the lensing signal is measured on the resampled blocks to estimate the variance of the measurement.
We compute the lensing signal for the different LRG samples (full sample, slow-growing, fast-growing), using a modified version of the athena\footnote{http://www.cosmostat.org/software/athena/} software, a 2d-tree code estimating second-order correlation functions from input galaxy catalogues.
	
\begin{figure}
	\centering
	\includegraphics{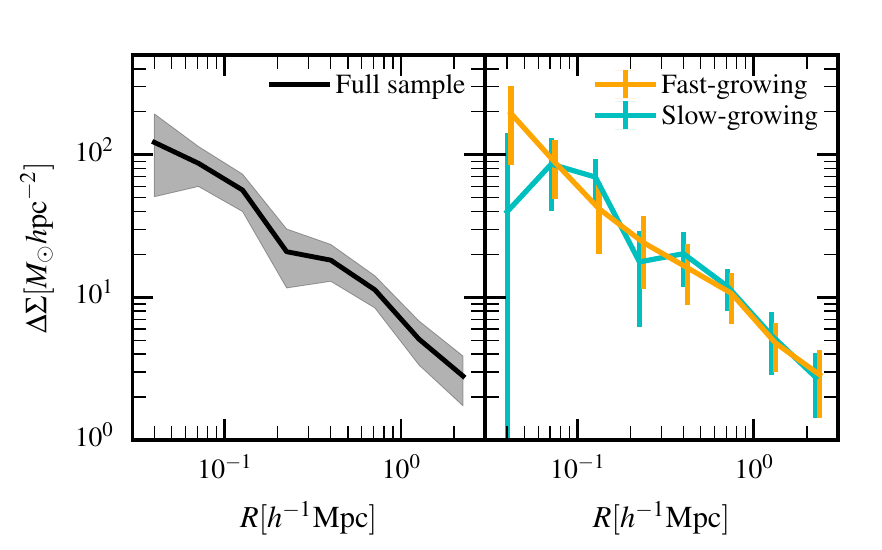}	
	\caption{ The left panel presents the lensing signal profile measured for the full LRG sample. In the right panel, results for fast- and slow-growing LRGs are shown in orange and cyan, respectively. Errors are computed using a block bootstrap on the data.
          }
	\label{fig:lensing_signal}
\end{figure}

Figure \ref{fig:lensing_signal} displays the excess surface mass density profile for the full LRG sample (\textit{left}), and the fast- and slow-growing LRG subsamples (\textit{right}). The error bars are estimated from the  diagonal terms of the covariance matrix obtained using the block bootstrap. 
Figure \ref{fig:lensing_signal} shows that the two LRG populations have similar excess surface mass density profiles, compatible within one sigma.
In the next section, we fit a halo model to these measurements to obtain and compare halo masses for fast- and slow-growing LRGs.

\section{The halo mass of LRGs}
\label{sec:halo_mass}

In order to extract halo masses from the measured lensing signal, we need to model the observed projected mass density. In the halo model formalism, dark matter  is assumed to be bound in haloes of different masses. In this view, the dark matter distribution surrounding the lens galaxy can be decomposed into three terms: the (sub)halo surrounding the galaxy (\textit{1h term}), the host halo if the galaxy is a satellite in a group or cluster (\textit{host term}), and the neighbouring haloes (\textit{2h term}). Adding the contribution from the stars in the galaxy, the excess surface mass density profile can be modeled as \citep[see also][]{cooray&sheth2002, gillis2013}:
\begin{equation}
\Delta\Sigma = \Delta\Sigma_{\mathrm{stars}} +  \Delta\Sigma_{\mathrm{1h}} + f_{\mathrm{sat}} \Delta\Sigma_{\mathrm{host}} +  \Delta\Sigma_{\mathrm{2h}}\mathrm{,}
\end{equation} 
where $\Delta\Sigma_{\mathrm{1h}} = (1 - f_{\mathrm{sat}})\Delta\Sigma_{\mathrm{1h}}^{\mathrm{cen}} + f_{\mathrm{sat}}\Delta\Sigma_{\mathrm{1h}}^{\mathrm{sub}}$. We model the two 1h terms (for centrals and satellites) with NFW density profiles parametrized by a mass $M_{\mathrm{halo}}$, but use different mass-concentration relations: the total mass selected relation from \citet{klypin2016} for centrals\footnote{We verify on the full-sample that using other $c-M$ relations \citep{dutton2014, shan2017} does not change our results.} and the relation from \citet{pastormira2011} for satellites to account for tidal stripping. We model the stars as a point source term with mass equal to the median stellar mass of the sample, and the host halo term as in \citet{niemiec2017}, with the difference that in this case the spatial distribution of satellite galaxies within their host haloes is unknown. We therefore assume that satellite galaxies follow the dark matter distribution, but with a lower concentration \citep{budzynski2012, wojtak2013}. We express $P(R_{\mathrm{s}})$, i.e. the probability for a satellite to be located at a distance $R_{\mathrm{s}}$ from the centre of its host cluster \citep[see equation 9 in][]{niemiec2017}, as a NFW profile with concentration $c_{\mathrm{sat}} = 2.5$\footnote{We tried fixing alternatively $c_{\mathrm{sat}} = 5$ and $7.5$ for the full sample and found no significant variation in the result, showing that with the current statistical uncertainties the halo masses are robust with respect to the choice of $c_{\mathrm{sat}}$.}. The 2h-term is as in \citet{niemiec2017}. All the quantities are expressed in comoving units.

The full model has three free parameters, the halo mass, $M_{\mathrm{halo}}$, the host mass, $M_{\mathrm{host}}$, and the satellite fraction, $f_{\mathrm{sat}}$. In our fiducial model, we reduce the number of free parameters by  fixing the value of the satellite fraction at $f_{\mathrm{sat}} = 0.1$ \citep{rodriguez-torres2016}.

\begin{figure*}
	\centering
	\includegraphics{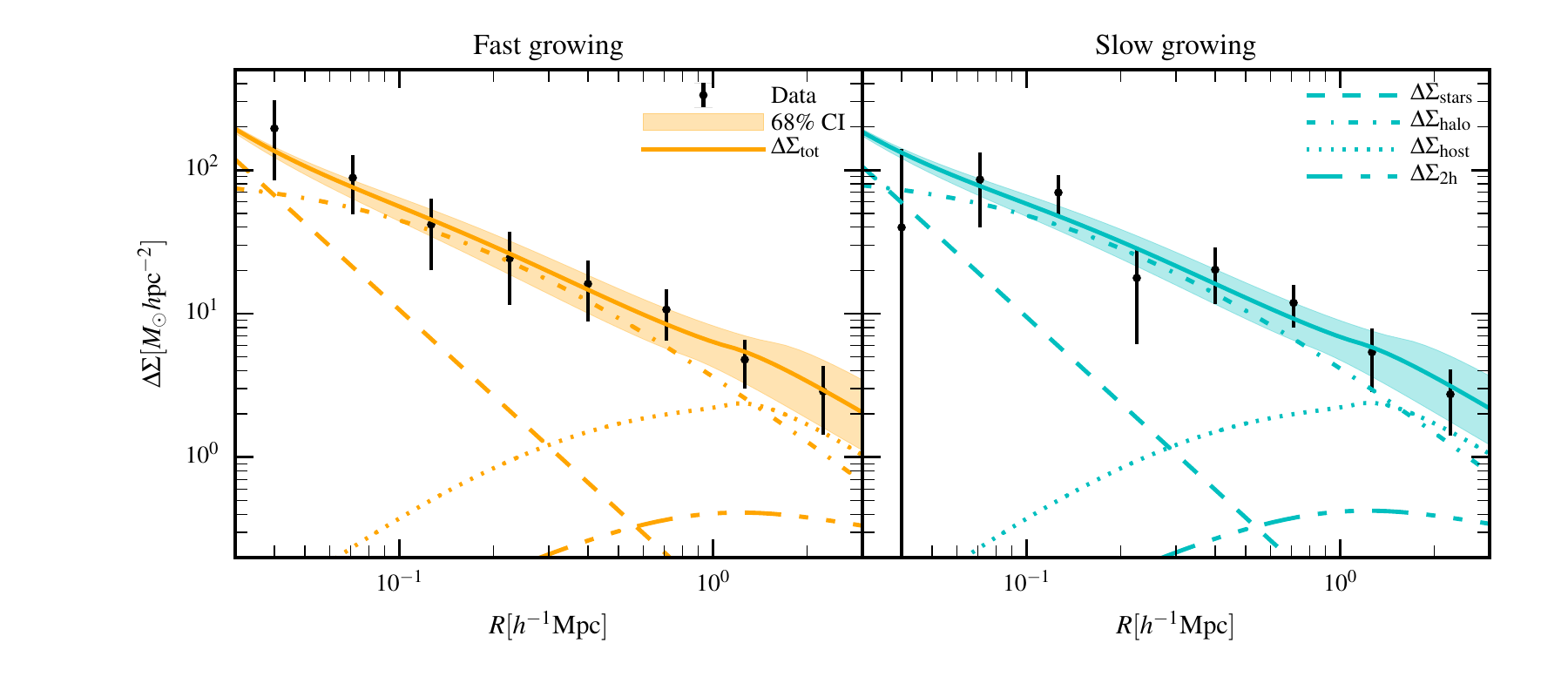}
	\caption{ Lensing signal for the fast- (\textit{left panel}) and slow-growing (\textit{right panel}) LRGs, with the best fit model (solid blue line) and $68\%$ confidence interval (blue surface). The four terms of the model are also shown.
          }
	\label{fig:lensing_wmodel}
\end{figure*}

We obtain the best-fit parameters and the credible intervals through a Markov Chain Monte Carlo (MCMC) method using emcee \citep{foreman-mackey2013} which is a Python implementation of an affine invariant MCMC ensemble sampler.
We define the likelihood $\mathcal{L}$ as:
\begin{equation}
	\mathcal{L} = \frac{1}{\sqrt{2\pi |C|}}\exp(-\frac{1}{2}(X_{\mathrm{obs}} - X_{\mathrm{mod}})^\mathrm{T}C^{-1}(X_{\mathrm{obs}} - X_{\mathrm{mod}}))
\end{equation}
where $X_{\mathrm{obs}}$ are the measurements in the 8 radial bins, $X_{\mathrm{mod}}$ the corresponding model predictions, and $C$ the covariance matrix from the block bootstrap. 
We assume flat and broad priors, such as: $M_{\mathrm{halo}} \in [10; 13.5]$ and $M_{\mathrm{host}} \in [13.5; 16]$.

\begin{figure*}
\centering
\begin{tabular}{ccc}
\subfloat[Full sample]{\includegraphics[height=5.cm]{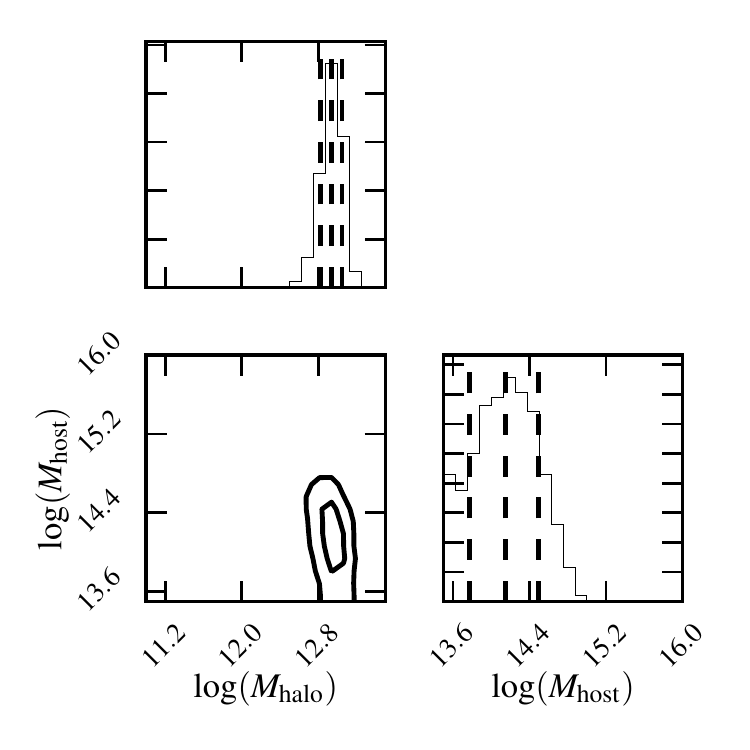}}   & \subfloat[Fast-growing]{\includegraphics[height=5.cm]{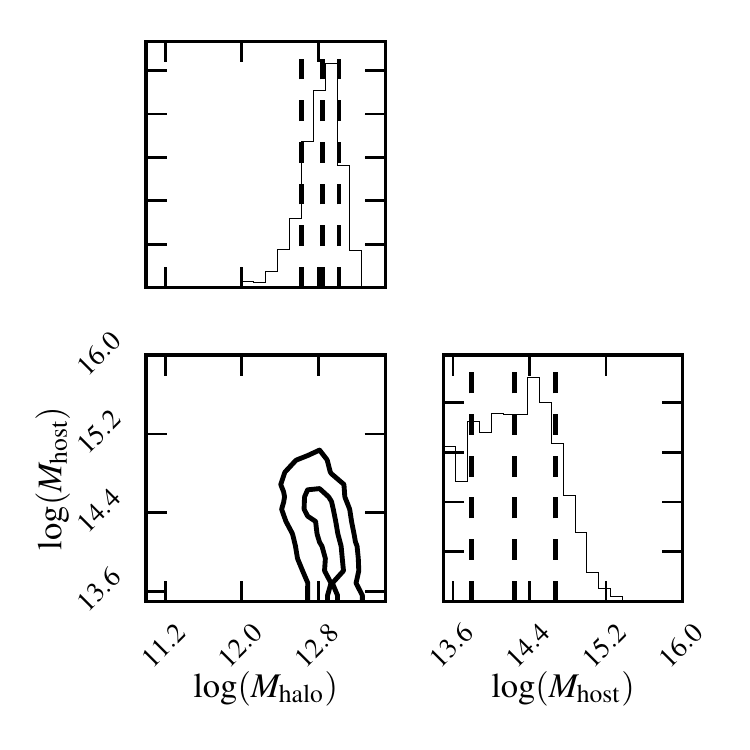}} & \subfloat[Slow-growing]{\includegraphics[height=5.cm]{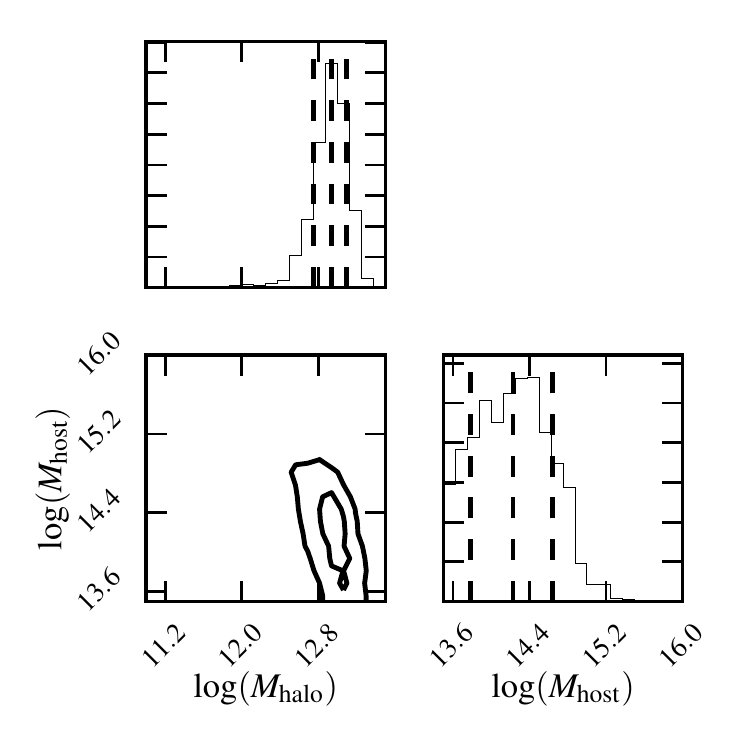}}\\
\end{tabular}
\caption{ Joint 2-dimensional and marginalized 1-dimensional posterior probability distributions for our two parameters $M_{\rm halo}$ and $M_{\rm host}$ for each of the three LRGs selections: full sample (\textit{left panel}), fast-growing (\textit{middle panel}) and slow-growing (\textit{left panel}). Masses are expressed in units of $h^{-1}\mathrm{M}_{\odot}$.}
\label{fig:corner_fiducial}
\end{figure*}

The best-fit models (solid lines) along with the $68\%$ credible intervals (shaded area) for the two samples are presented in Figure \ref{fig:lensing_wmodel}, where the lensing measurements are also provided for comparison.
We plot the four terms of the model in different dashed and dotted lines.  We list in Table \ref{tab:lenses} the  best-fit parameters and 68\% credible intervals for the fiducial model, described by the 16th and 84th percentile of the posterior probability distribution. We also show in Figure \ref{fig:corner_fiducial}  the joint 2D and marginalized 1D posterior probability distributions for the two parameters $M_{\mathrm{halo}}$ and $M_{\mathrm{host}}$.

We find that the two LRG populations live in haloes of very similar mass: $\log\left(M_{\rm halo}^{\rm fast}/h^{-1}\mathrm{M}_{\odot}\right) = 12.85^{+0.16}_{-0.26}$ and  $\log\left(M_{\rm halo}^{\rm slow}/h^{-1}\mathrm{M}_{\odot}\right) =12.92^{+0.16}_{-0.22}$, respectively. 
For the host halo term, we obtain a typical mass $M_{\mathrm{host}} \sim 10^{14}\mathrm{M}_{\odot}$, which is in broad agreement with the BOSS CMASS clustering results such as \citet{rodriguez-torres2016}.
However, we note that the halo mass obtained in the 1h term is lower than predicted by clustering analyses, which has been thoroughly discussed in \citet{leauthaud2016}, but which does not affect our conclusions.

\begin{table*}
\centering
\begin{tabular}{c | c c c c c c c}
			&	$N_{\rm lenses}$	&	$<\log{M_{*}/\rm{M}_{\odot}}>$	&	$<z_{\rm l}>$	&	$\log(M_{\rm halo}/h^{-1}\rm{M}_{\odot})$	&	$\log(M_{\rm host}/h^{-1}\rm{M}_{\odot})$		\\
\hline
Full sample	&	6972				&	11.66					&	0.55			&	$12.94^{+0.11}_{-0.12}$				&	$14.15^{+0.34}_{-0.37}$					\\

Fast-growing	&	3356				&	11.68					&	0.55			&	$12.85^{+0.17}_{-0.22}$				&	$14.24^{+0.42}_{-0.45}$					\\

Slow-growing	&	3616				&	11.63					&	0.55			&	$12.94^{+0.16}_{-0.19}$				&	$14.23^{+0.41}_{-0.44}$					\\

\hline

\end{tabular}
\caption{Description of the lens samples and best-fit parameters for the fiducial model.}
\label{tab:lenses}
\end{table*}

As an alternative fit, we also allow for the satellite fraction $f_{\mathrm{sat}}$ to vary between 0 and 1. As expected, the $f_{\mathrm{sat}}$ and $M_{\mathrm{host}}$ parameters are degenerate, as a low fraction of satellites living in massive host haloes will give a similar lensing signal as a larger fraction living in less massive hosts. We still obtain values of $f_{\mathrm{sat}}$ consistent with 0.1 although not very tightly constrained: $f_{\mathrm{sat}}^{\mathrm{fast}} = 0.25^{+0.27}_{-0.18}$ and $f_{\mathrm{sat}}^{\mathrm{slow}} = 0.17^{+0.28}_{-0.13}$. It is important to note that we still obtain consistent values for the halo mass within reasonable error bars: $\log\left(M_{\mathrm{halo}}^{\mathrm{fast}}/h^{-1}\mathrm{M}_{\odot}\right) = 12.68^{+0.27}_{-0.53}$ and $\log\left(M_{\mathrm{halo}}^{\mathrm{slow}}/h^{-1}\mathrm{M}_{\odot}\right) = 12.82^{+0.20}_{-0.39}$.

We have also tried leaving the halo concentration parameter free, in addition  to the halo and host mass, but the current uncertainties on our measurements prevent us from constraining the three parameters simultaneously.

\section{Discussion and conclusion}
\label{sec:conclusion}

The galaxy-galaxy weak lensing measurements presented in this work for  fast- and slow-growing LRGs show that the two galaxy populations present very similar excess surface mass density profiles. Fitting a halo model to these measurements gives a similar mean halo mass for the samples, with $M_{\mathrm{halo}} \sim 10^{13} \mathrm{M}_{\odot}$. 
This result, together with the clustering amplitude difference and the similarity in the stellar mass distribution presented in \citet{montero-dorta2017}, suggest the existence of galaxy assembly bias.

However, using the bias-halo mass relation from \citet{tinker2010}, we estimate that a $10\%$ bias difference at $b \simeq 2$, as reported in \citet{montero-dorta2017} among fast- and slow-growing LRGs, corresponds to a halo mass difference of only $\sim 0.15$ dex. This small mass difference could not be detected with the current size of error bars, and we therefore cannot definitively conclude to the evidence of galaxy assembly bias. In order to measure such a weak difference in halo mass, we should decrease our error bars by a factor of $\sim 5$, which corresponds to an increase of the effective area used in the lensing measurement of a factor of $\sim 25$. Having such shear catalogues covering $\sim 5000 \deg^2$ should soon be possible with surveys such as DES, which would allow to settle the question.

\vspace{0.3cm}

\begingroup
    \fontsize{8pt}{0.5}\selectfont
 Based on observations obtained with MegaPrime/MegaCam, a joint project of CFHT and CEA/DAPNIA, at the Canada-France-Hawaii Telescope (CFHT), which is operated by the National Research Council (NRC) of Canada, the Institut National des Science de l'Univers of the Centre National de la Recherche Scientifique (CNRS) of France, and the University of Hawaii. The Brazilian partnership on CFHT is managed by the Laborat\'orio Nacional de Astrof\'isica (LNA). We thank the support of the Laborat\'orio Interinstitucional de e-Astronomia (LIneA). We thank the CFHTLenS team. This work was granted access to the HPC resources of Aix-Marseille Universit\'{e} financed by the project Equip@Meso (ANR-10-EQPX-29-01) of the program "Investissements d'Avenir" supervised by the Agence Nationale pour la Recherche. EP is supported by MINECO grants AYA2016-77846- P and AYA2014-57490-P and Junta de Andaluc\'ia P12- FQM-2828. ADMD, FP and SRT are supported by MINECO grant AYA2014-60641-C2-1-P. AMD acknowledges support from the Funda\c{c}\~ao de Amparo \`a Pesquisa do Estado de S\~ao Paulo (FAPESP), through the grant 2016/23567--4.
\endgroup

\bibliographystyle{mn2e}
\bibliography{/Users/annaniem/Biblio/lensing.bib}

\begin{thebibliography}{43}
\expandafter\ifx\csname natexlab\endcsname\relax\def\natexlab#1{#1}\fi

\bibitem[{{Alam} {et~al}\mbox{.}(2015){Alam}, {Albareti}, {Allende Prieto},
  {Anders}, {Anderson}, {Anderton}, {Andrews}, {Armengaud}, {Aubourg},
  {Bailey}, \& et~al.}]{alam2015}
{Alam} S. {et~al.}, 2015, \apjs, 219, 12

\bibitem[{{Ben{\'{\i}}tez}(2000)}]{benitez2000}
{Ben{\'{\i}}tez} N., 2000, \apj, 536, 571

\bibitem[{{Brainerd}, {Blandford} \& {Smail}(1996){Brainerd}, {Blandford}, \&
  {Smail}}]{brainerd1996}
{Brainerd} T.~G., {Blandford} R.~D., {Smail} I., 1996, \apj, 466, 623

\bibitem[{{Budzynski} {et~al}\mbox{.}(2012){Budzynski}, {Koposov}, {McCarthy},
  {McGee}, \& {Belokurov}}]{budzynski2012}
{Budzynski} J.~M., {Koposov} S.~E., {McCarthy} I.~G., {McGee} S.~L.,
  {Belokurov} V., 2012, \mnras, 423, 104

\bibitem[{{Bundy} {et~al}\mbox{.}(2015){Bundy}, {Leauthaud}, {Saito}, {Bolton},
  {Lin}, {Maraston}, {Nichol}, {Schneider}, {Thomas}, \& {Wake}}]{bundy2015}
{Bundy} K. {et~al.}, 2015, \apjs, 221, 15

\bibitem[{{Busch} \& {White}(2017)}]{busch&white2017}
{Busch} P., {White} S.~D.~M., 2017, \mnras, 470, 4767

\bibitem[{{Cid Fernandes} {et~al}\mbox{.}(2005){Cid Fernandes}, {Mateus},
  {Sodr{\'e}}, {Stasi{\'n}ska}, \& {Gomes}}]{cid_fernandes2005}
{Cid Fernandes} R., {Mateus} A., {Sodr{\'e}} L., {Stasi{\'n}ska} G., {Gomes}
  J.~M., 2005, \mnras, 358, 363

\bibitem[{{Coil} {et~al}\mbox{.}(2017){Coil}, {Mendez}, {Eisenstein}, \&
  {Moustakas}}]{coil2017}
{Coil} A.~L., {Mendez} A.~J., {Eisenstein} D.~J., {Moustakas} J., 2017, \apj,
  838, 87

\bibitem[{{Coil} {et~al}\mbox{.}(2008){Coil}, {Newman}, {Croton}, {Cooper},
  {Davis}, {Faber}, {Gerke}, {Koo}, {Padmanabhan}, {Wechsler}, \&
  {Weiner}}]{coil2008}
{Coil} A.~L. {et~al.}, 2008, \apj, 672, 153

\bibitem[{{Cooray} \& {Sheth}(2002)}]{cooray&sheth2002}
{Cooray} A., {Sheth} R., 2002, Physics Reports, 372, 1

\bibitem[{{Dutton} \& {Macci{\`o}}(2014)}]{dutton2014}
{Dutton} A.~A., {Macci{\`o}} A.~V., 2014, \mnras, 441, 3359

\bibitem[{{Dvornik} {et~al}\mbox{.}(2017){Dvornik}, {Cacciato}, {Kuijken},
  {Viola}, {Hoekstra}, {Nakajima}, {van Uitert}, {Brouwer}, {Choi}, {Erben},
  {Fenech Conti}, {Farrow}, {Herbonnet}, {Heymans}, {Hildebrandt}, {Hopkins},
  {McFarland}, {Norberg}, {Schneider}, {Sif{\'o}n}, {Valentijn}, \&
  {Wang}}]{dvornik2017}
{Dvornik} A. {et~al.}, 2017, \mnras, 468, 3251

\bibitem[{{Foreman-Mackey} {et~al}\mbox{.}(2013){Foreman-Mackey}, {Hogg},
  {Lang}, \& {Goodman}}]{foreman-mackey2013}
{Foreman-Mackey} D., {Hogg} D.~W., {Lang} D., {Goodman} J., 2013, \pasp, 125,
  306

\bibitem[{{Gao}, {Springel} \& {White}(2005){Gao}, {Springel}, \&
  {White}}]{gao2005}
{Gao} L., {Springel} V., {White} S.~D.~M., 2005, \mnras, 363, L66

\bibitem[{{Gao} \& {White}(2007)}]{gao&white2007}
{Gao} L., {White} S.~D.~M., 2007, \mnras, 377, L5

\bibitem[{{Gillis} {et~al}\mbox{.}(2013){Gillis}, {Hudson}, {Erben}, {Heymans},
  {Hildebrandt}, {Hoekstra}, {Kitching}, {Mellier}, {Miller}, {van Waerbeke},
  {Bonnett}, {Coupon}, {Fu}, {Hilbert}, {Rowe}, {Schrabback}, {Semboloni}, {van
  Uitert}, \& {Velander}}]{gillis2013}
{Gillis} B.~R. {et~al.}, 2013, \mnras, 431, 1439

\bibitem[{{Guo} {et~al}\mbox{.}(2013){Guo}, {Zehavi}, {Zheng}, {Weinberg},
  {Berlind}, {Blanton}, {Chen}, {Eisenstein}, {Ho}, {Kazin}, {Manera},
  {Maraston}, {McBride}, {Nuza}, {Padmanabhan}, {Parejko}, {Percival}, {Ross},
  {Ross}, {Samushia}, {S{\'a}nchez}, {Schlegel}, {Schneider}, {Skibba},
  {Swanson}, {Tinker}, {Tojeiro}, {Wake}, {White}, {Bahcall}, {Bizyaev},
  {Brewington}, {Bundy}, {da Costa}, {Ebelke}, {Malanushenko}, {Malanushenko},
  {Oravetz}, {Rossi}, {Simmons}, {Snedden}, {Streblyanska}, \&
  {Thomas}}]{guo2013}
{Guo} H. {et~al.}, 2013, \apj, 767, 122

\bibitem[{{Heymans} {et~al}\mbox{.}(2012){Heymans}, {Van Waerbeke}, {Miller},
  {Erben}, {Hildebrandt}, {Hoekstra}, {Kitching}, {Mellier}, {Simon},
  {Bonnett}, {Coupon}, {Fu}, {Harnois D{\'e}raps}, {Hudson}, {Kilbinger},
  {Kuijken}, {Rowe}, {Schrabback}, {Semboloni}, {van Uitert}, {Vafaei}, \&
  {Velander}}]{heymans2012}
{Heymans} C. {et~al.}, 2012, \mnras, 427, 146

\bibitem[{{Hildebrandt} {et~al}\mbox{.}(2012){Hildebrandt}, {Erben}, {Kuijken},
  {van Waerbeke}, {Heymans}, {Coupon}, {Benjamin}, {Bonnett}, {Fu}, {Hoekstra},
  {Kitching}, {Mellier}, {Miller}, {Velander}, {Hudson}, {Rowe}, {Schrabback},
  {Semboloni}, \& {Ben{\'{\i}}tez}}]{hildebrandt2012}
{Hildebrandt} H. {et~al.}, 2012, \mnras, 421, 2355

\bibitem[{{Klypin} {et~al}\mbox{.}(2016){Klypin}, {Yepes}, {Gottl{\"o}ber},
  {Prada}, \& {He{\ss}}}]{klypin2016}
{Klypin} A., {Yepes} G., {Gottl{\"o}ber} S., {Prada} F., {He{\ss}} S., 2016,
  \mnras, 457, 4340

\bibitem[{{Leauthaud} {et~al}\mbox{.}(2017){Leauthaud}, {Saito}, {Hilbert},
  {Barreira}, {More}, {White}, {Alam}, {Behroozi}, {Bundy}, {Coupon}, {Erben},
  {Heymans}, {Hildebrandt}, {Mandelbaum}, {Miller}, {Moraes}, {Pereira},
  {Rodr{\'{\i}}guez-Torres}, {Schmidt}, {Shan}, {Viel}, \&
  {Villaescusa-Navarro}}]{leauthaud2016}
{Leauthaud} A. {et~al.}, 2017, \mnras, 467, 3024

\bibitem[{{Lin} {et~al}\mbox{.}(2016){Lin}, {Mandelbaum}, {Huang}, {Huang},
  {Dalal}, {Diemer}, {Jian}, \& {Kravtsov}}]{lin2016}
{Lin} Y.-T., {Mandelbaum} R., {Huang} Y.-H., {Huang} H.-J., {Dalal} N.,
  {Diemer} B., {Jian} H.-Y., {Kravtsov} A., 2016, \apj, 819, 119

\bibitem[{{Miller} {et~al}\mbox{.}(2013){Miller}, {Heymans}, {Kitching}, {van
  Waerbeke}, {Erben}, {Hildebrandt}, {Hoekstra}, {Mellier}, {Rowe}, {Coupon},
  {Dietrich}, {Fu}, {Harnois-D{\'e}raps}, {Hudson}, {Kilbinger}, {Kuijken},
  {Schrabback}, {Semboloni}, {Vafaei}, \& {Velander}}]{miller2013}
{Miller} L. {et~al.}, 2013, \mnras, 429, 2858

\bibitem[{{Miyatake} {et~al}\mbox{.}(2016){Miyatake}, {More}, {Takada},
  {Spergel}, {Mandelbaum}, {Rykoff}, \& {Rozo}}]{miyatake2016}
{Miyatake} H., {More} S., {Takada} M., {Spergel} D.~N., {Mandelbaum} R.,
  {Rykoff} E.~S., {Rozo} E., 2016, Physical Review Letters, 116, 041301

\bibitem[{{Montero-Dorta} {et~al}\mbox{.}(2017){Montero-Dorta}, {P{\'e}rez},
  {Prada}, {Rodr{\'{\i}}guez-Torres}, {Favole}, {Klypin}, {Cid Fernandes},
  {Gonz{\'a}lez Delgado}, {Dom{\'{\i}}nguez}, {Bolton}, {Garc{\'{\i}}a-Benito},
  {Jullo}, \& {Niemiec}}]{montero-dorta2017}
{Montero-Dorta} A.~D. {et~al.}, 2017, \apjl, 848, L2

\bibitem[{{Moraes} {et~al}\mbox{.}(2014){Moraes}, {Kneib}, {Leauthaud},
  {Makler}, {Van Waerbeke}, {Bundy}, {Erben}, {Heymans}, {Hildebrandt},
  {Miller}, {Shan}, {Woods}, {Charbonnier}, \& {Pereira}}]{moraes2014}
{Moraes} B. {et~al.}, 2014, in Revista Mexicana de Astronomia y Astrofisica
  Conference Series, Vol.~44, pp. 202--203

\bibitem[{{More} {et~al}\mbox{.}(2016){More}, {Miyatake}, {Takada}, {Diemer},
  {Kravtsov}, {Dalal}, {More}, {Murata}, {Mandelbaum}, {Rozo}, {Rykoff},
  {Oguri}, \& {Spergel}}]{more2016}
{More} S. {et~al.}, 2016, \apj, 825, 39

\bibitem[{{Niemiec} {et~al}\mbox{.}(2017){Niemiec}, {Jullo}, {Limousin},
  {Giocoli}, {Erben}, {Hildebrant}, {Kneib}, {Leauthaud}, {Makler}, {Moraes},
  {Pereira}, {Shan}, {Rozo}, {Rykoff}, \& {Van Waerbeke}}]{niemiec2017}
{Niemiec} A. {et~al.}, 2017, \mnras, 471, 1153

\bibitem[{{Pastor Mira} {et~al}\mbox{.}(2011){Pastor Mira}, {Hilbert},
  {Hartlap}, \& {Schneider}}]{pastormira2011}
{Pastor Mira} E., {Hilbert} S., {Hartlap} J., {Schneider} P., 2011, \aap, 531,
  A169

\bibitem[{{Planck Collaboration} {et~al}\mbox{.}(2014){Planck Collaboration},
  {Ade}, {Aghanim}, {Armitage-Caplan}, {Arnaud}, {Ashdown}, {Atrio-Barandela},
  {Aumont}, {Baccigalupi}, {Banday}, \& et~al.}]{planck2014}
{Planck Collaboration} {et~al.}, 2014, \aap, 571, A16

\bibitem[{{Rodr{\'{\i}}guez-Torres}
  {et~al}\mbox{.}(2016){Rodr{\'{\i}}guez-Torres}, {Chuang}, {Prada}, {Guo},
  {Klypin}, {Behroozi}, {Hahn}, {Comparat}, {Yepes}, {Montero-Dorta},
  {Brownstein}, {Maraston}, {McBride}, {Tinker}, {Gottl{\"o}ber}, {Favole},
  {Shu}, {Kitaura}, {Bolton}, {Scoccimarro}, {Samushia}, {Schlegel},
  {Schneider}, \& {Thomas}}]{rodriguez-torres2016}
{Rodr{\'{\i}}guez-Torres} S.~A. {et~al.}, 2016, \mnras, 460, 1173

\bibitem[{{Shan} {et~al}\mbox{.}(2017){Shan}, {Kneib}, {Li}, {Comparat},
  {Erben}, {Makler}, {Moraes}, {Van Waerbeke}, {Taylor}, {Charbonnier}, \&
  {Pereira}}]{shan2017}
{Shan} H. {et~al.}, 2017, \apj, 840, 104

\bibitem[{{Sunayama} {et~al}\mbox{.}(2016){Sunayama}, {Hearin}, {Padmanabhan},
  \& {Leauthaud}}]{sunayama2016}
{Sunayama} T., {Hearin} A.~P., {Padmanabhan} N., {Leauthaud} A., 2016, \mnras,
  458, 1510

\bibitem[{{Tinker} {et~al}\mbox{.}(2010){Tinker}, {Robertson}, {Kravtsov},
  {Klypin}, {Warren}, {Yepes}, \& {Gottl{\"o}ber}}]{tinker2010}
{Tinker} J.~L., {Robertson} B.~E., {Kravtsov} A.~V., {Klypin} A., {Warren}
  M.~S., {Yepes} G., {Gottl{\"o}ber} S., 2010, \apj, 724, 878

\bibitem[{{van Uitert} {et~al}\mbox{.}(2016){van Uitert}, {Cacciato},
  {Hoekstra}, {Brouwer}, {Sif{\'o}n}, {Viola}, {Baldry}, {Bland-Hawthorn},
  {Brough}, {Brown}, {Choi}, {Driver}, {Erben}, {Heymans}, {Hildebrandt},
  {Joachimi}, {Kuijken}, {Liske}, {Loveday}, {McFarland}, {Miller}, {Nakajima},
  {Peacock}, {Radovich}, {Robotham}, {Schneider}, {Sikkema}, {Taylor}, \&
  {Verdoes Kleijn}}]{vanuitert2016}
{van Uitert} E. {et~al.}, 2016, \mnras, 459, 3251

\bibitem[{{Velander} {et~al}\mbox{.}(2014){Velander}, {van Uitert}, {Hoekstra},
  {Coupon}, {Erben}, {Heymans}, {Hildebrandt}, {Kitching}, {Mellier}, {Miller},
  {Van Waerbeke}, {Bonnett}, {Fu}, {Giodini}, {Hudson}, {Kuijken}, {Rowe},
  {Schrabback}, \& {Semboloni}}]{velander2014}
{Velander} M. {et~al.}, 2014, \mnras, 437, 2111

\bibitem[{{Wang} {et~al}\mbox{.}(2011){Wang}, {Navarro}, {Frenk}, {White},
  {Springel}, {Jenkins}, {Helmi}, {Ludlow}, \& {Vogelsberger}}]{wang2011}
{Wang} J. {et~al.}, 2011, \mnras, 413, 1373

\bibitem[{{Wechsler} {et~al}\mbox{.}(2006){Wechsler}, {Zentner}, {Bullock},
  {Kravtsov}, \& {Allgood}}]{wechsler2006}
{Wechsler} R.~H., {Zentner} A.~R., {Bullock} J.~S., {Kravtsov} A.~V., {Allgood}
  B., 2006, \apj, 652, 71

\bibitem[{{Wojtak} \& {Mamon}(2013)}]{wojtak2013}
{Wojtak} R., {Mamon} G.~A., 2013, \mnras, 428, 2407

\bibitem[{{Yang}, {Mo} \& {van den Bosch}(2006){Yang}, {Mo}, \& {van den
  Bosch}}]{yang&vandenbosch2006}
{Yang} X., {Mo} H.~J., {van den Bosch} F.~C., 2006, \apjl, 638, L55

\bibitem[{{Zehavi} {et~al}\mbox{.}(2011){Zehavi}, {Zheng}, {Weinberg},
  {Blanton}, {Bahcall}, {Berlind}, {Brinkmann}, {Frieman}, {Gunn}, {Lupton},
  {Nichol}, {Percival}, {Schneider}, {Skibba}, {Strauss}, {Tegmark}, \&
  {York}}]{zehavi2011}
{Zehavi} I. {et~al.}, 2011, \apj, 736, 59

\bibitem[{{Zentner}, {Hearin} \& {van den Bosch}(2014){Zentner}, {Hearin}, \&
  {van den Bosch}}]{zentner2014}
{Zentner} A.~R., {Hearin} A.~P., {van den Bosch} F.~C., 2014, \mnras, 443, 3044

\bibitem[{{Zu} {et~al}\mbox{.}(2017){Zu}, {Mandelbaum}, {Simet}, {Rozo}, \&
  {Rykoff}}]{zu2017b}
{Zu} Y., {Mandelbaum} R., {Simet} M., {Rozo} E., {Rykoff} E.~S., 2017, \mnras,
  470, 551

\end{thebibliography}

\label{lastpage}
\end{document}